\documentclass[letterpaper,twoside,english]{article}
\usepackage[T1]{fontenc}
\usepackage[latin1]{inputenc}
\usepackage{graphicx}
\usepackage[authoryear]{natbib}

\makeatletter

\usepackage{latexsym,times}
\usepackage[intlimits]{amsmath}
\usepackage{amsfonts,amssymb}
\DeclareSymbolFontAlphabet{\mathbb}{AMSb}
\usepackage[mathscr]{eucal}
\usepackage[textstyle,cdot,squaren,thickspace,thickqspace]{SIunits}
\usepackage{fancyvrb}
\usepackage{url}
\usepackage{xspace}
\usepackage{psfrag}
\usepackage[ps2pdf]{thumbpdf}              
\usepackage[%
bookmarks=true,
bookmarksnumbered=true,
hypertexnames=false,
pdfborder={0 0 0},
ps2pdf]{hyperref}%
\hypersetup{
  pdfauthor   = {can ozan tan <tanc@cns.bu.edu>},
  pdftitle    = {Artificial neural networks - method comp.},
  pdfsubject  = {ANNs},
  pdfkeywords = {ANNs, backprop, ARTMAP},
  pdfcreator  = {LaTeX with hyperref package},
  pdfproducer = {dvips and ps2pdf}
}
\usepackage[letterpaper,top=1in,bottom=1in,hmargin={1in,1in}]{geometry}
\numberwithin{equation}{section}
\usepackage{float}
\usepackage[bf,float,subfigure]{caption2}
\setcaptionmargin{0.5in}
\usepackage[sf]{subfigure}

\DeclareRobustCommand\em
{\@nomath\em \ifdim \fontdimen\@ne\font >\z@
  \upshape \else \slshape \fi}

\usepackage{babel}
\makeatother
\begin{document}
\date{}

\title{Statistical Predictive Models in Ecology: Comparison of Performances
and Assessment of Applicability}

\author{Can Ozan Tan$^{1,*}$, Uygar \"{O}zesmi$^{2}$, Meryem Beklioglu$^{3}$,
Esra Per$^{4}$, Bahtiyar Kurt$^{4}$ }

\maketitle
\noindent $^{1}$Canakkale Onsekiz Mart University, Department of
Biological Sciences, Terzioglu Kampusu, Canakkale, Turkey TR-17020.

\noindent $^{2}$Erciyes University, Department of Environmental Engineering,
Kayseri, Turkey TR-38039.

\noindent $^{3}$Middle East Technical University, Department of Biological
Sciences, Ankara, Turkey TR-06531.

\noindent $^{4}$Doga Dernegi P.O. 640, Yenisehir, Ankara TR-06445,
Turkey

\noindent $^{*}$Corresponding author: Boston University, Department
of Cognitive and Neural Systems, 677 Beacon Street \#201 , Boston
MA, 02215. Tel: +1 617 353 6741. Fax: +1 617 353 7755. e-mail: tanc@cns.bu.edu

\newpage

\flushleft

\begin{abstract}

Ecological systems are governed by complex interactions which are mainly nonlinear. 
In order to capture the inherent complexity and nonlinearity of ecological, and in 
general biological systems, statistical models recently gained popularity. However, 
although these models, particularly connectionist approaches such as multilayered 
backpropagation networks, are commonly applied as predictive models in ecology to 
a wide variety of ecosystems and questions, there are no studies to date aiming to 
assess the performance, both in terms of data fitting and generalizability, and 
applicability of statistical models in ecology. Our aim is hence to provide an 
overview for nature of the wide range of the data sets and predictive variables, 
from both aquatic and terrestrial ecosystems with different scales of time-dependent 
dynamics, and the applicability and robustness of predictive modeling methods on such 
data sets by comparing different statistical modeling approaches. The models used in 
this study range from predicting the occurrence of submerged plants in shallow lakes 
to predicting nest occurrence of bird species from environmental variables and satellite 
images. The methods considered include k-nearest neighbor (k-NN), linear and quadratic 
discriminant analysis (LDA and QDA), generalized linear models (GLM) feedforward
multilayer backpropagation networks and pseudo-supervised network ARTMAP.

Our results show that the predictive performances of the models on training data 
could be misleading, and one should consider the predictive performance of a given 
model on an independent test set for assessing its predictive power. 
Moreover, our results suggest that for ecosystems involving time-dependent
dynamics and periodicities whose frequency are possibly less than
the time scale of the data considered, GLM and connectionist neural
network models appear to be most suitable and robust, provided that a predictive 
variable reflecting these time-dependent dynamics included in the model either implicitly 
or explicitly. For spatial data, which does not include any
time-dependence comparable to the time scale covered by the data,
on the other hand, neighborhood based methods such as k-NN and ARTMAP
proved to be more robust than other methods considered in this study.
In addition, for predictive modeling purposes, first a suitable, computationally 
inexpensive method should be applied to the problem at hand a good predictive 
performance of which would render the computational
cost and efforts associated with complex variants unnecessary. 

\end{abstract}
\newpage

\section{Introduction}

\label{sec:introduction}

Ecological, and in general biological, systems' dynamics are often
governed with nonlinear interactions of environmental factors. Environmental
variables interact with systems in such a complex way that the whole
system achieves a broader functionality that cannot be deduced by
considering individual environmental factors. The goal of the ecological
studies has been to gain insight into this functionality and complexity
by observing individual factors affecting the system in question.
System dynamics, and the influence of individual factors to these
dynamics as a whole, has been of primary concern not only for theoretical
considerations, but also management and conservation practices. One
natural choice to achieve this purpose is to abstract the system and
interactions inherent in it by mathematical and statistical models.

Ecological modeling studies have traditionally been concentrated on
the use of \char`\"{}box-and-arrow\char`\"{} type differential/difference
equation models \citep{ross76,jorgensen76,lassiter:kearns77}. However,
while these models provide valuable insight into the particular ecosystem
being studied, because of their strong dependence on the parameters
specific to that system, these models are prone to be criticized as
being \char`\"{}case-specific\char`\"{}. To achieve generalizability
and thus to gain insight into the ecosystems in general, statistical
models have been one of the main practices (e.g. \citealt{tan:smeins96,maron:lill2004,gutierrez:etal2005,tan:beklioglu2005,ozesmi:etalinpress2}).
This is especially true for statistical classification and pattern
recognition models which fundamentally aim to characterize a combination
of variables and their measurements which lead to particular state(s)
of the ecosystem. 

On the other hand, traditional classification, and in general statistical,
models require strong assumptions about the distribution of the underlying
observations and/or the system characteristics \citep{hastie:etal01}.
To overcome this issue, nonlinear and/or non-parametric models to
model ecosystems on the basis of finite-size observations have become
recently popular.
Of particular importance in this respect is the use of connectionist
artificial neural network-based approaches such as generalized linear
models and feedforward multilayer backpropagation networks. These
models are able to capture the nonlinear interactions and complexity
of the ecosystem without the need for any assumptions about the distribution
of the observations. However, the main problem associated with these
approaches has been their \char`\"{}off-line\char`\"{} iterative nature
\citep{bishop95,ripley96,hastie:etal01} and the computational costs
associated with these models, compared to traditional approaches \citep{hastie:etal01}.

The latter problem can be overcome by considering the use of simpler
models where they attain a fairly good performance. The former problem,
on the other hand, being off-line, constitutes to the main issue,
especially for management practices. Each time a new observation is
made, these models require to be retrained on the entire data set
in order to include the new observation. Considering the computational
cost associated with artificial neural network based approaches, this
fact renders these models less practical from a management point of
view. However, though unfamiliar to ecological and biological modelers,
several other families of neural network models have been developed
since the early 70's, including ART and ARTMAP family of models \citep{grossberg1976a,grossberg1976b,carpenter:etal1991a,carpenter:etal1992}.
These unsupervised and pseudo-supervised, self organizing maps have
the advantage of being real-time models, and thus being able to incorporate
new observations easily. In addition, the computational costs associated
with ART-family of models are comparably small compared to connectionist
neural network models, while their performance is reasonably better.

To assess the performance of a model, several studies showed the importance
of independent test sets \citep{ozesmi:etalinpress,tan:beklioglu2005,tan:beklioglu:subm}.
Although a given model, particularly neural networks \citep{hornik:etal1989},
can perform arbitrarily well on training data set, the actual goal
for statistical models is to achieve a high generalizability at the
same time. From this point of view, a given model should minimize
the training error and at the same time maximize the generalizability
(so-called minimax problem). In that respect, while connectionist
neural network schemes are \char`\"{}universal approximators\char`\"{}
of the training data set \citep{hornik:etal1989}, there has been
several techniques developed to maximize the predictive performance
of a given model simultaneously on both training and test sets, such
as cross-validation \citep{bishop95,hastie:etal01}.

Recently, as mentioned before, the use of statistical techniques gained
popularity to explain and predict the outcome of ecosystem processes,
such as occurrence of multiple stable states \citep{tan:smeins96,tan:beklioglu2005,tan:beklioglu:subm},
and habitat selection or distribution of species \citep{baran:etal96,lek:etal96,ozesmis99}.
Predictive modeling attempts in ecology have used several methods
ranging from regression \citep{sawchik:etal2003,gutierrez:etal2005}
to discriminant analysis \citep{joy:death2003,maron:lill2004}, and
from generalized additive \citep{Seoane:etal2003b,seoane:etal2004b,dunk:etal2004}
and linear models \citep{meggs:etal2004,tan:beklioglu2005} to multilayer
feedforward backpropagation networks \citep{scardi96,tan:smeins96,ozesmis99}
and time series analysis \citep{heegaard2002}. However, considering
above mentioned arguments, it is necessary to assess predictive statistical
models and algorithms in terms of their performance on and suitability
for a given ecological problem, which to our knowledge, has not been
done yet. Although there are extensive studies aiming to compare statistical
models for artificial data sets (e.g. STATLOG project; \citealt{michie:etal1994}),
it is obvious that the ecological data differs considerably than 
artificial data sets since former data is considerably more prone
to observational and/or measurement noise, and the ecological interactions
are inherently more complex and nonlinear. In addition, while all
the variables and factors leading to system dynamics in artificial
systems can be known and/or controlled, such is not true for the ecological
systems. As such, a comparative study to asses the performance and
at the same time suitability of statistical modeling techniques is
required for the ecological modeling studies to avoid blind applications
and ill allocation of time and effort.

Our aim is here to provide an overview for nature of the wide range
of the data sets and predictive variables and their use in predictive ecological models by comparing different
statistical modeling approaches. 
To that end, we studied 6 different data sets from both aquatic and
terrestrial ecosystems using 6 different type of statistical models.
The models included traditional methods (k-nearest neighbor: k-NN,
and linear and quadratic discriminant analysis: LDA and QDA), connectionist
neural network-type models (generalized linear models: GLM, feedforward
multilayer backpropagation networks) and pseudo-supervised ARTMAP.
Some of these data sets used in this
paper were modeled before using one or more of the particular algorithms
and discussed extensively in several publications \citep{ozesmi96,ozesmis99,ozesmi:etalinpress,ozesmi:etalinpress2,per2003,bahtiyar04,tan:beklioglu2005,tan:beklioglu:subm} and as such, only a broad summary of these data sets are provided among
with the relevant citations.

\section{Data Sets}

\subsection{Data Set 1: Lakes Eymir and Mogan}

\label{sec:eymir:mogan}

Lake Eymir is a small shallow lake located in Central Anatolia. The
upstream Lake Mogan empties into Lake Eymir at the southwest corner,
forming the main inflow \citep{beklioglu:etal03}. 

Data used in the model for Lake Eymir and Lake Mogan were collected
between 1997 and 2002 \citep{beklioglu:etal03,tan02}. A total of
91 data points from Lake Eymir were used for fitting the models and
the a total of 43 data points from the upstream Lake Mogan were used
as an independent test set after fitting the models, to determine
ability of the models to generalize.

Concentrations of total phosphorus (TP, $\mu$gl$^{-1}$), suspended
solids (SS, mg l$^{-1}$), and chlorophyll-a (chl-a, $\mu$gl$^{-1}$),
Secchi disk transparency (Secchi, cm) and water levels (WL, meter above
sea level) were used as predictive (independent) variables. Dependent
variable was a binary index of submerged plant occurrence. It was
suggested that the impacts of submerged plants become apparent on
ecology of shallow lakes when both the plant volume infested (PVI\%, 
\citealp{canfield:etal84} and the coverage exceeded 30\% \citep{sondergaard:moss97}.
The formulation of the dichotomous dependent variable was thus

\begin{equation}
\mathrm{Y}=\left\{ \begin{array}{ll}
1 & \mathrm{if}\, C_{P}>0.3\,\mathrm{and}\, V_{P}>0.3\\
0 & \mathrm{otherwise}\end{array}\right\} \label{eq:glm:output}\end{equation}

\noindent where $Y$ is the binary index to show presence/absence
of submerged vegetation; (0: absent; 1: present); $C_{P}\in[0.0,1.0]$ is the plant cover and
$V_{P}\in[0.0,1.0]$ is the PVI. Detailed description of the study sites, data
collection as well as a through discussion of the ecological state
of these lakes are given elsewhere \citep{tan02,beklioglu:etal03,beklioglu:etal04,tan:beklioglu2005}.

\subsection{Data Set 2: Central Anatolian Shallow Lakes}

\label{sec:5lakes}

Data comes from five lakes (Lakes Bey\c{s}ehir, I\c{s}\i kl\i, Marmara,
Mogan and Uluabat), which vary in size and depth and are located from
southern to northern Anatolia, Turkey. These lakes were selected to
model the impacts of fish biomass, hydrology and morphology on submerged
plant development. 

Data used to fit (equivalently to train) and test the models span
several years, ranging from 19 years as in the case of Bey\c{s}ehir
to only two years as in the case of Lake Mogan. A total of 541 data
points for all the lakes were obtained from the literature and pooled
together for fitting the models. Pooled data were further randomly
split to two sets of 440 and 101 data points which constitute our
final training and test data, respectively. The validation test set
was not included in the training, and was reserved for assessment
of the model performance on an independent data set (referred hereafter
as validation data for set 2). In addition, the data gathered from
Lake Mogan, consisting of 24 data points, was not included either
in the training data or in the validation data sets, and was reserved
for a second test set to asses the generalizability of the fitted
models. This set corresponds to a system spatially and temporally
distinct from the data used to train the model (referred hereafter
as independent data for set 2).

Models consisted of 5 predictive variables, which included the ratio
of carp (\textit{Cyprinus carpio}) biomass to total fish biomass (carp
ratio), amplitude of the intra-annual water level fluctuation defined
as the difference between yearly maximum and minimum water depth,
morphology index, calculated for each lake as the monthly ratio $Z_{mean}/Z_{max}$
averaged over whole period spanned by the data, where $Z_{mean}$
is the mean depth and $Z_{max}$ is the maximum depth. Last two predictive variables were the z-score
of water level, and period index, which was simply a sine transformation
of the Julian date of the corresponding data point. Occurrence of
submerged plants, assessed by equation {[}\ref{eq:glm:output}{]}
was used as dependent variable. In some rare cases, where there were
no available quantitative data about the submerged plant coverage
and/or volume in the lakes, but qualitative information is provided,
we used the latter to designate the dichotomous dependent variable. 

For a detailed description of the predictive and dependent variables,
readers are referred to \cite{tan:beklioglu:subm}. A detailed description
of the study sites, data collection as well as a through discussion
of the ecological states of these lakes are given elsewhere \citep{tan02,beklioglu:etal04,tan:beklioglu:subm}.

\subsection{Data Sets 3 and 4: Nest Occurrence and Breeding Success and of Red-Winged
Blackbird}

\label{sec:robertson:erie}

Data are collected from two marshes in 1995 and 1996 , in Sandusky
Bay on Lake Erie, Ohio, USA and collected in 1969 and 1970 from other two
located in Connecticut, in the northeastern USA. The Lake Erie marshes
were Stubble Patch and Darr. Data from these marshes included
habitat variables and nest occurrence of red-winged blackbird (\textit{Agelaius
phoeniceus}) \citep{ozesmi96}. A detailed description of the study
sites and data can be found in \cite{ozesmi96}. Two marshes in Connecticut
were Clarkes Pond and All Saints Marsh. The data from these marshes
included habitat information and breeding success of the same
species \citep{robertson72}. A detailed description of the sites
and data collection can be found in \cite{robertson72}.

We built two separate sets of models using different dependent variables:
nest occurrence and breeding success models. The nest occurrence model
was fit using 1995 data from the Lake Erie wetlands Stubble Patch
and Darr (data set 3). The predictive variables were vegetation durability
based on an ordinate scale between 0 and 100 \citep{ozesmi:mitsch97},
stem density (number of stems m$^{-2}$), stem height (cm above water),
distance to open water (m), distance to edge (m), and water depth
(cm). The dependent variable was a binary index of nest occurrence. 

In the breeding success model, the predictive variables were vegetation
durability based on an ordinate scale between 0 and 100 \citep{ozesmi:mitsch97},
nest height (cm), distance to open water (m), distance to edge (m)
and water depth (cm). A binary index of whether or not any nestling
fledged was the dependent variable.

Data from Clarkes Pond collected in 1969 - 1970 is used for fitting
the breeding success models (data set 4). In total 294 data points
were available to fit the models. The final model was tested using
the data from All Saints Marsh from 1969 ($n=101$) and from 1970
($n=130$) as independent tests to assess the generalizability. Trained
models were also tested on the Lake Erie wetlands data from Darr and
Stubble Patch in 1995 and 1996. Similarly, nest occurrence models,
trained on the Lake Erie data, were tested on the Connecticut wetlands
data from Clarkes Pond and All Saints in 1969 and 1970. Because one
set of the models were developed to predict breeding success and the
other nest occurrence, for this research we assumed that a high probability
of nest occurrence corresponds to a high probability of breeding success
and vice versa. In addition, since the Connecticut wetland variables
did not include stem density, the average value of stem density from
the Lake Erie wetlands was used when testing Connecticut wetlands
data on the Lake Erie model. Note also that nest height was available
for the Connecticut wetlands while stem height was available for
the Lake Erie wetlands, and these variables were used interchangeably.
Stem heights were about 50 cm higher than nest heights on average.
The assumption was made that stem height and nest height were correlated.
Thus when the models were tested, nest heights were used as predictive
variable instead of stem heights in the nest occurrence model and
vice versa for the breeding success model. In addition to the nest
occurrence data from Darr and Stubble Patch, the information on breeding
success from Darr and Stubble Patch in 1996 together with the habitat
variables of vegetation durability, nest height, distance to open
water, distance to edge, and water depth under the nest was used as
test data for the Clarkes Pond breeding success model.

\subsection{Data Set 5: Breeding Presence of Three Bird Species}

\label{sec:bahtiyar}

Data set 5 collected in spring 2001 and spring 2002 in southeastern
Turkey to predict the presence or absence of breeding bird species
depending on the environmental variables. The data gathered included
three bird species: woodchat shrike (\textit{Lanius senator;} Linnaeus,
1758), short-toed lark (\textit{Calandrella brachydactyla;} Leisler,
1814), and olivaceous warbler (\textit{Hippolais pallida;} Ehrenberg,
1833). Data sets were consisting of $N_{L}=548$, $N_{C}=490$ and
$N_{H}=598$ data points for \textit{L. senator}, \textit{C. brachydactyla}
and \textit{H. pallida}, respectively.

There were 12 predictive variables. 6 of them were satellite image
data from LANDSAT corresponding to the bands TM1-TM5 and TM7. Each
of these bands are generally implicated in reflecting a different
attribute of the vegetation cover and land geography \citep{bahtiyar04}.
Remaining predictive variables included annual mean temperature ($^{o}$C),
annual mean humidity (\%), distance to nearest road (m), distance
to nearest water source (m), the height of the sampling point from
sea level, and a vegetation index, which was a dimensionless categorical
variable indicating the type of the vegetation with respect to its
height, ranging between 1 - 9, 1 being the shortest vegetation type
\citep{bahtiyar04}. The dependent variable was binary indicating
the breeding presence or absence of bird species in question.

We split the data set for each species into two by randomly selecting
half of the data corresponding to each output category for each species
and sparing those selected data as test sets (independent tests),
while using the other half for training the models. Hence, three sets
of each model considered in this study were built separately and fit
on half of the data set for each species and then tested on the other
half of the data, which were not used during training, for that particular
species. Detailed description of the study sites and data collection
among with a through discussion of the biology of the bird species
can be found in \citet{bahtiyar04}.

\subsection{Data Set 6: Habitat Selection of Bird Species in Central Anatolia}

This particular data set concerned the habitat preferences of 9 bird
species in Central Anatolia, namely great reed warbler \textit{(Acrocephalus
arundinaceus} Linnaeus, 1978), skylark (\textit{Alauda arvensis} Linnaeus,
1758), short-toed lark (\textit{Calandrella brachydactyla} Leisler,
1814), lesser short-toed lark (\textit{Calandrella rufescens} Vieillot,
1820), marsh harrier (\textit{Circus aeruginosus} Linnaeus, 1758),
calandra lark (\textit{Melanocorypha calandra} Linnaeus, 1766), corn
bunting (\textit{Milaria calandra} Linnaeus, 1758), yellow wagtail
(\textit{Motacilla flava} Linnaeus, 1758) and isabellina wheatear
(\textit{Oenanthe isabellina} Isabellina Temminck, 1829). The data
has been collected from Sultan marshes and Tuzla Lake in Central Anatolia
during 2002. 

In field studies, the presence (1) or absence (0) of each bird species on
a particular spot were recorded along with the environmental variables, and
were used as dependent variable for models. 
There were 12 predictive variables. These were vegetation
index, which was a categorical variable with 23 categories \citep{per2003},
percent vegetation cover (\%), stem height (cm), water depth (cm),
grazing, which was a semantic variable with 4 categories ranging from
0 (none) to 3 (extensive), and 6 satellite imagery bands. The images
used were obtained from LANDSAT satellite, and the bands used as independent
variables were TM1-5 and TM7. Sample size for each species is given
in Table \ref{tab:esra}. Detailed description of the satellite images,
and the properties of the bands used here, as well a a description
of the study sites and the biology of the bird species considered
can be found in \cite{per2003}. 

For this data set 9 separate sets of models have been built for each
species. Five of them (sets 1 - 5) were fit to the data collected
in Sultan marshes for each species and then tested on the data collected
in Lake Tuzla for that species (independent tests). Remaining 4 set
of models (sets 6 - 9) were fit to the half of the data randomly split
from data collected in Lake Tuzla for each species separately, and
models were then tested on the remaining half (validation).

\section{Statistical Methods}

In this section, an overview of the pattern recognition and classification
models used for this study are reviewed. It is not meant to be exhaustive,
and interested readers are referred to the authoritative references
in relevant sections. A short description of the specific implementation
details for the current study is provided at the end of the section.

\subsection{Preprocessing}

\label{sec:standardization}

Often, predictive variables in a given model are of different units
which are not compatible with each other. For instance a model may
include both a distance measure given in meters as well as the concentration
of a particular nutrient given in $\mu$gl$^{-1}$ (e.g. data sets
1 and 2; sections \ref{sec:eymir:mogan} and \ref{sec:5lakes}) as
a predictive variable. Or it may include a dimensionless variable
among several others with compatible units (e.g. section \ref{sec:robertson:erie}).
Similarly, some predictive variables could be given in compatible,
or even same units but spanning different ranges of values. For example
the data set from lakes Eymir and Mogan (data set 1; section \ref{sec:eymir:mogan})
include chlorophyll-a concentration as well as total phosphorus concentration,
both given in $\mu$gl$^{-1}$. But while the former spans a value
range between $[1,38]$, the latter variable spans a value range of
$[54,532]$. Such incompatibilities are known to reduce the stability
and performance of statistical models as the initial randomization
of model parameters, or the weights, will not be effective if the
predictive variables are on different scales \citep{bishop95,hastie:etal01}.
In such cases, to ensure the stability and convergence of the model
to a solution, it is necessary to standardize the input variables
to a consistent dimensionless interval.

One commonly employed standardization scheme is to linearly transform
the independent variables to mean of zero and units of standard deviation,
also known as z-score transformation \citep{fisher1970}, such that
the $i$th value of the $k$th predictive variable is transformed
as:

\begin{equation}
z_{i,k}=\frac{\mu_{k}-x_{i,k}}{\sigma_{k}}\end{equation}

\noindent where $\mu_{k}$ and $\sigma_{k}$ are the mean and standard
deviation of the $k$th variable.

If standard deviations differ substantially among variables (as is
usually the case if binary or categorical variables are included in
the input space), it is preferable to linearly transform the values
of all predictive variables to lie in the range of $\pm0.5$ \citep{bishop95,goodman96:manual}.
Note that in this case, the mean will not be zero unless the mean
of the raw predictor was centered between its minimum and maximum
values.

One other approach to standardize the input space is to use hypercube
transformation such that after transforming, input variables lie in
a space ${\mathcal{{C}}}^{P}\in[0,1]^{P}$ where $P$ is the number
of input variables:

\begin{equation}
x_{i,k}^{\mathcal{{C}}}=\frac{x_{i,k}-\min(x_{i,k})}{\max(x_{i,k})-\min(x_{i,k})}\end{equation}

\noindent where $x_{i,k}^{\mathcal{{C}}}\in[0,1]^{P}$ is the transformed
data point $x_{i,k}$. In addition to standardization, hypercube transformation
also provides a strictly bounded input space, and thus it is of theoretical
importance for classification, and in general statistical modeling
techniques, especially for ART family of models (section \ref{sec:artmap}).
A review of these theoretical considerations is beyond the scope of
this paper, and interested readers are referred to \cite{grossberg1988},
\cite{carpenter:etal1991c} and \citet{kosko1992}. In all cases,
after fitting the model, the variables can then be inversely transformed
for predictive purposes. 

All data used in this study were z-score transformed before feeding
into the models except for ARTMAP, for which the input data were hypercube
transformed.

\subsection{Model Fitting}

\subsubsection{Traditional Classification Models}

\label{sec:traditional}

For comparative purposes, we used k-nearest neighbor (k-NN), linear
and quadratic discriminant (LDA and QDA) methods to classify our data
points according to the output classes (dependent variables). k-NN
method has been considered as a benchmark classification method, if
one considers only the training data. This method uses those observations
in the training set $\mathcal{{T}}$ closest in the input space to
$x$ to form $\hat{Y}$. More specifically,

\begin{equation}
\hat{Y}=\frac{1}{k}\sum_{x_{i}\in N_{k}(x)}y_{i}\end{equation}

\noindent where $N_{k}(x)$ is the neighborhood of $x$ defined by
the $k$ closest points $x_{i}$ in the training sample. It is clear
that when the neighborhood $k$ is considered to be $k=1$, k-NN method
potentially can reach to minimum classification error on the training
set. Note that in this case the error on test set is expected to be
quite high. Thus, k-NN method with a neighborhood size $k=2$ is employed
in this study as a benchmark of training set performance.

LDA and QDA techniques are models based on the class densities of
the output categories. In other words, they enable one to infer the
posterior probabilities of the output categories based on the data
observed, using Bayes theorem:

\begin{equation}
\mathrm{P}(G=k|X=x)=\frac{f_{k}(x)\pi_{k}}{\sum_{l=1}^{K}f_{l}(x)\pi_{l}}\end{equation}

\noindent where $f_{k}(x)$ is the class-conditional density of $X$
in class $G=k$, and $\pi_{k}$ is the prior probability of class
$k$ with $\sum_{k=1}^{K}\pi_{k}=1$. LDA and QDA assume Gaussian
distribution for class densities. Fundamentally, for two category
cases (as in our case), and assuming that the covariances $\Sigma_{k}$
of the class densities are equal, linear discriminant function is
given as

\begin{equation}
\delta_{K}=x^{T}\Sigma^{-1}\mu_{k}-\frac{1}{2}\mu_{k}^{T}\Sigma^{-1}\mu_{k}+\log\pi_{k}\end{equation}

\noindent where the parameters of the Gaussian distributions are estimated
from the data as

\begin{eqnarray}
\hat{\pi}_{k} & = & \frac{N_{k}}{N}\\
\hat{\mu}_{k} & = & \frac{\sum_{g_{i}=k}x_{i}}{N_{k}}\\
\hat{\Sigma} & = & \frac{\sum_{k=1}^{K}\sum_{g_{i}=k}(x_{i}-\hat{\mu}_{k})(x_{i}-\hat{\mu}_{k})^{T}}{(N-K)}\end{eqnarray}

\noindent where $N_{k}$ is the number of class-$k$ observations.
An equivalent decision rule is given as $G(x)=\mathrm{arg}\max_{k}\delta_{k}(x)$.
If the equality assumption of class covariances $\Sigma_{k}$ does
not hold, we obtain quadratic discriminant function

\begin{equation}
\delta_{k}(x)=-\frac{1}{2}\log|\Sigma_{k}|-\frac{1}{2}(x-\mu_{k})^{T}\Sigma_{k}^{-1}(x-\mu_{k})+\log\pi_{k}\end{equation}

\noindent with an equivalent decision boundary between each pairs
of classes $k$ and $l$ described by a quadratic equation $\{ x:\delta_{k}(x)=\delta_{l}(x)\}$.

Both LDA and QDA are shown to perform astonishingly well on large
and diverse set of classification tasks, and both techniques are widely
used in various research areas \citep{michie:etal1994}. Thus, we
included these two models as potential benchmarks to compare the performances
of other methods against, assuming that the data considered in this
study are distributed following a Gaussian distribution. Note that
discriminant analyses, both linear and quadratic, strictly require
that the underlying data are distributed as a Gaussian. A more in-depth
discussion of these two methods, among with k-NN method, can be found
in \cite{hastie:etal01}

\subsubsection{Generalized Linear Models and Feedforward Multilayer Backpropagation
Networks}

\label{sec:glm}

A general linear model is similar to classical multiple regression
analysis such that the explanatory variables, $\mathbf{X}$, multiplied
by weights, $\beta$, obtained by statistical estimation, are summed
together for a score, $\mathbf{Y}$:

\begin{equation}
\mathbf{Y}_{i}=\sum_{j}\beta_{j}x_{j}+\epsilon_{i}=\mathbf{X\beta}+\epsilon_{i}\label{eq:glm:main}\end{equation}

\noindent Setting 

\noindent \begin{equation}
\nu_{i}=\mathbf{X}\beta+\epsilon_{i}\label{eq:glm:nu}\end{equation}

\noindent a general liner model is analogous to multiple regression
models with the estimated values being transformed through a nonlinear
signal function (sigmoid, in our case)
For simple linear prediction with continuous (analog) dependent variables,
$\nu$ might have intrinsic meaning. For binary outcome events, as
in our cases, however, a link function that monotonically constrains
the output prediction to lie between 0 and 1 is required. An asymmetric
logistic function, ranging from 0 to 1, is a common choice, as it
is analogous to the probability that a given pattern is associated
with a particular output class:

\begin{equation}
\mathbf{Y}_i = P_{i}=\frac{e^{\nu_{i}}}{1+e^{\nu_{i}}}\label{eq:glm:out}\end{equation}

For training and performance assessment purposes, a score threshold
$P_{t}$ is required to assign the predicted probability to one of
the output classes such that if the output prediction is above $P_{t}$
it is classified as 1 and to 0 if $P_{i}<P_{t}$. An obvious choice
for binary outcome events is $P_{t}=0.5$.

As with the case of multiple regression, general linear models (GLM)
are equivalent to connectionist neural networks without any hidden
layers. GLM are designed to emphasize the linear combination of predictive
variables in explaining the dependent variable(s), as they do not
include any processes for nonlinearly transforming the input space.
However, ecological, or more generally biological systems inherently
include nonlinearities \citep{may76,scheffer:etal93:ASS}, which may
severely limit the performance of linear models \citep{ozesmi:etalinpress,tan:beklioglu:subm}.
For that reason, artificial neural networks, particularly feedforward
multilayer backpropagation networks, which are designed to capture
the nonlinear interactions in the input space, are favored in recent
years \citep{baran:etal96,lek:etal96,lek:guegan99,scardi96,scardi01,ozesmis99}.
On the other hand, artificial neural networks are computationally
expensive compared to linear models. In some cases they offer only
a little, if any, improvement over generalized linear models \citep{bishop95,goodman96:manual,tan:beklioglu2005}.
GLMs have been successfully applied to several data sets in ecology (e.g. \citealp{meggs:etal2004,tan:beklioglu2005}).

Connectionist neural networks are among the most applied and well
known class of supervised statistical models. These networks are composed
of an input layer, an arbitrary number of hidden layers with arbitrary
number of hidden units in each layer, and an output layer. The network
is commonly (but not necessarily) fully connected, meaning that each
node in a given layer $l_{j}$ is connected to all of the nodes in
the next layer $l_{j+1}$. These networks are also feedforward, such
that the 'information' flows from input layer to hidden layer(s) to
output layer in only forward direction. Feedforward neural networks
are formally defined as follows:

Let $A^{r}$ be the set of all affine functions from $R^{r}$ to $R$,
that is the set of all functions in the form $A(x)=w\cdot x+b$. For
any measurable function $G(\cdot)$ mapping $R$ to $R$ and $r\in N$,
$\Sigma\Pi^{r}(G)$ is the class of functions

\begin{equation}
\left\{ f:R^{r}\rightarrow R:f(x)=\sum_{j=1}^{q}\beta_{j}\cdot\prod_{k=1}^{l_{j}}G(A_{jk}(x))\right\} ,x\in R^{r},\beta_{j}\in R,A_{jk}\in A^{r},l_{j}\in N,q=1,2,...\end{equation}

\noindent where the product term disappears for networks without hidden
layer, and the equation reduces to that of generalized linear model.

The function $G(\cdot)$ is the activation function, or link function,
equivalent to that of output units in general linear model. Again,
for binary outcome event, an asymmetric logistic activation function,
ranging from 0 to 1, is suitable for output unit. However, a symmetric
logistic activation function, ranging between $\pm$0.5, is generally
suggested for the hidden units since a symmetric function is shown
to enhance the stability of the weights of hidden units during backpropagation
of errors \citep{rumelhart86,unbehauen:cichocki96}. \label{sec:optimization}

Multilayer feedforward backpropagation networks have been successfully
applied to a diverse set of data ranging from phytoplankton production
\citep{scardi96,scardi01} to community changes based on climatic
inputs \citep{tan:smeins96}, and to relationships of different species
to habitat variables \citep{baran:etal96,reyjol:etal01,ozesmis99,tan:beklioglu:subm}.
Model fitting procedure for feedforward neural networks, as well as
for general linear models are discussed in detail in \citet{rumelhart86},
\cite{bishop95} and \cite{ripley96}. Implementational details of
GLM and backpropagation models for the data sets used in this study
are given in \cite{tan:beklioglu2005}, \cite{tan:beklioglu:subm},
\cite{ozesmis99}, \cite{bahtiyar04}, and \cite{per2003}.

\subsubsection{ARTMAP}

\label{sec:artmap}

\begin{figure}
\begin{center}\includegraphics[%
  height=0.70\textheight,
  keepaspectratio]{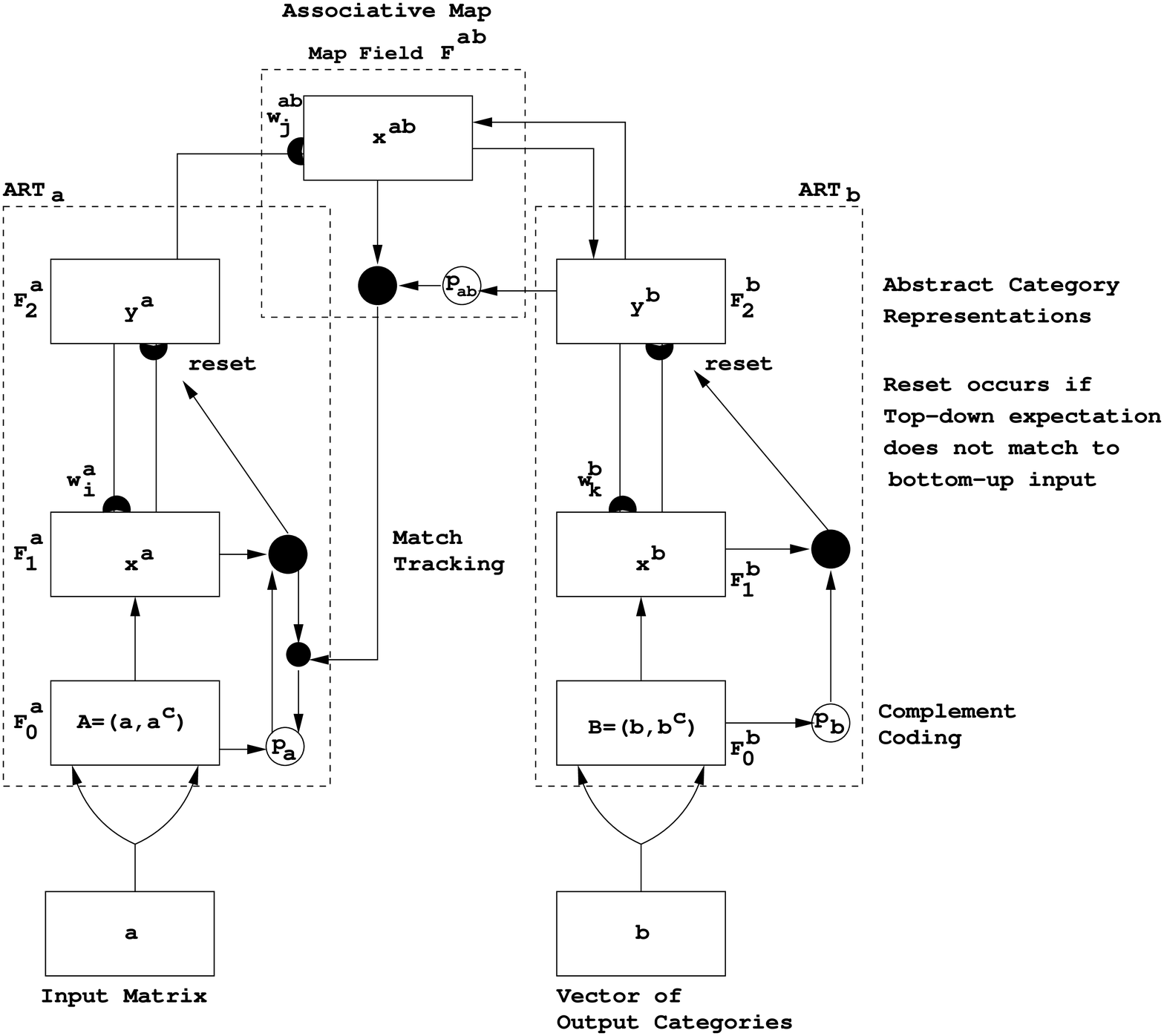}\caption{}\label{fig:fart}\end{center}
\end{figure}

A schematic presentation of generic ARTMAP model is provided in Figure
\ref{fig:fart}. Briefly, ARTMAP models consist basically of two so-called
ART modules, which are fundamentally self-organizing maps \citep{carpenter:etal1991a},
one for input space and one for output space\textbf{.} Learning occurs
separately for each ART module independently, whenever an expected
category matches to presented input pattern (i.e., current combination
of predictive variables), or a novel input pattern is encountered.
These modules are linked by an associative learning network and an
internal controller that ensures autonomous system operation in real
time. Thus, ARTMAP models represent a \char`\"{}pseudo-supervised\char`\"{}
learning method \citep{carpenter:etal1991a}. The controller is designed
to create a minimal number of ART recognition categories (committed
nodes; that is, abstract representations of combinations of input
vectors) for the input space, or \char`\"{}hidden units\char`\"{}
analogous to backpropagation networks, needed to meet an accuracy
criteria, which is given by the so-called \char`\"{}vigilance\char`\"{}
parameter $\rho$ \citep{carpenter:etal1992}. ARTMAP algorithm fundamentally
works by increasing the vigilance parameter of the input ART module
by the minimal amount needed to correct a predictive error at ART
module of the output classes.

There are several variants of ART modules \citep{carpenter:grossberg1990,carpenter:etal1991b,carpenter:etal1991c}.
Here, we used fuzzy ART modules, which are developed for pattern recognition
models with continuous and/or categorical input space \citep{carpenter:etal1991c,carpenter:etal1992}.
Fuzzy ART and fuzzy ARTMAP models use fuzzy logic operators \citep{kosko1992}
for category choices and match criteria, as well as for learning in
the model operation.

Shortly, each ART system contains an input field $F_{0}$, a $F_{1}$
field receiving bottom-up signals from $F_{0}$ and top-down input
from $F_{2}$, the latter of which represents the active category
representations\textbf{.} So-called \textit{complement coding} \citep{carpenter:etal1992}
should be employed before feeding the input vectors to fuzzy ART modules.
Fundamentally by complement coding,
it is meant that an $M$-dimensional input matrix $\mathbf{a}$ is
coded and fed to the model as an $2M$-dimensional matrix $[\mathbf{a},\mathbf{a}^{c}]$,
where $a_{i}^{c}=(1-a_{i})$. 
Theoretical considerations for this requirement are discussed in detail
in \cite{carpenter:etal1992}.

At each $F_{2}$ category node, there is a weight associated with
that node, which are initially set to 1. Each weight $w_{ji}$ is
monotonically increasing with time and hence its convergence to a
limit is guaranteed \citep{carpenter:etal1991a,carpenter:etal1992}.
Fuzzy ART dynamics depend on a choice parameter $\alpha>0$, a learning
rate $\beta\in[0,1]$, and a maximum vigilance parameter $\rho_{\max} \in[0,1]$.
For each given input pattern and $j$th node of $F_{2}$ layer, the
choice function $T_{j}$ is defined by

\begin{equation}
T_{J}{(\mathbf{I})}=\frac{|\mathbf{I}\wedge\mathbf{w}_{j}|}{\alpha+|\mathbf{w}_{j}|}\end{equation}

\noindent where $\wedge$ is the fuzzy AND operator is equivalent
to component-wise $\min$ operator \citep{kosko1992}, $|\cdot|$
is the Euclidean norm, and $\mathbf{w}_{j}=(w_{j1}\cdots w_{jM})$.
The system makes a category choice when at most one $F_{2}$ node
can become at a given time, and the category choice is given as $T_{J}=\max\{ T_{j}:j=1\dots N\}$.
In a choice system, the activity of a given node at $F_{1}$ layer
is given as $\mathbf{x}=\mathbf{I}$ if $F_{2}$ node is inactive
and $\mathbf{x}=\mathbf{I}\wedge\mathbf{w}_{J}$ if $J$th $F_{2}$
node is selected. So-called 'resonance' occurs in the ART module if

\begin{equation}
\frac{|\mathbf{I}\wedge\mathbf{w}_{J}|}{|\mathbf{I}|}\geq\rho\label{eq:resonance}\end{equation}

\noindent and reset occurs otherwise \citep{carpenter:etal1991c,carpenter:etal1992}.
If reset occurs, the value of the choice function $T_{J}$ is set
to 0, and a new index $J$ is chosen. The search process continues
until the chosen $J$ satisfies the resonance criterion (equation
\ref{eq:resonance}). Once search ends and resonance occurs, the weight
vector $\mathbf{w}_{J}$ is updated by

\begin{equation}
\mathbf{w}_{J}^{(\mathrm{new})}=\beta\left(\mathbf{I}\wedge\mathbf{w}_{J}^{(\mathrm{old})}\right)+(1-\beta)\mathbf{w}_{J}^{(\mathrm{old})}.\end{equation}

As briefly mentioned above, fuzzy ARTMAP model consists of two fuzzy
ART modules, one for input and one for target vectors linked by an
associative learning network and an internal controller (Figure \ref{fig:fart}). When a prediction
by ART$_{a}$ module, which receives the input vectors, is disconfirmed
at ART$_{b}$ module, receiving target, or output, vector, inhibition
of map field activation induces the match tracking process, which
raises the ART$_{a}$ vigilance $\rho_{a}$ to just above the $F_{1}^{a}$
so that the activation of $F_{0}^{a}$ matches the reset criterion
(i.e., decreased just to miss the match criterion given by equation
\ref{eq:resonance}). This triggers an ART$_{a}$ search process which
leads to activation of either an ART$_{a}$ category that correctly
predict $b$ at match field, or to a new node which has not used before
(that is, either an already formed category that predicts $b$ is
selected, or a new category is created). Abstract category representations
formed at the end of a training session are termed as 'committed nodes'
and are analogous to the units in the hidden layer in connectionist
(i.e., multilayer feedforward backpropagation) networks.

A detailed review of the theory and operation of ARTMAP and fuzzy
ARTMAP models is beyond the scope of this study, and interested readers
are referred to \cite{carpenter:etal1991a}, \cite{carpenter:etal1991b}
and \cite{carpenter:etal1992}. \citet{carpenter:etal1991c} also
provides a geometric interpretation for ART algorithm. Although new
to ecology, ART and ARTMAP theory has been developed since early 70's,
and the reader is referred to \cite{cohen:grossberg1983} and \cite{grossberg1988}
for theoretical considerations. A compact review of implementational
issues can be found in \cite{carpenter2003}.

\subsection{Model Assessment and Validation}

\subsubsection{Performance criteria}

To asses the performance of statistical models, a score (error) is
employed which is to be maximized (minimized). For models with continuous-valued
dependent variables, a commonly employed choice is the least squares
error:

\begin{equation}
\min\mathrm{LSE}=\min|\mathbf{D}-\mathbf{Y}|\end{equation}

\noindent where $\mathbf{D}$ and $\mathbf{Y}$ are actual and predicted
values of the dependent variable, respectively. For binary or categorical
outcome events, however, a cross-entropy measure as a score measure
is preferred \citep{goodman96:manual,ripley96,hastie:etal01}:

\begin{equation}
C=\sum_{j,c}d_{j,c}\log_{2}(y_{j,c})+(1-d_{j,c})\log_{2}(1-y_{j,c})\label{eq:cindex}\end{equation}

\noindent where $d_{j,c}$ is the actual activation of output unit
and $y_{j,c}$ is the predicted activation. Cross-entropy, as calculated
by equation \ref{eq:cindex}, is approximately equal to the area under
receiver-operator curve \citep{goodman96:manual,ripley96}, and it
is also equivalent to log likelihood. Hence, maximizing equation \ref{eq:cindex}
is equivalent to maximizing the likelihood of the model estimate \citep{hastie:etal01}.
Note that by definition of ROC curve, c-index measures the ratio of
'hit' rate to 'false-alarm' rate. Thus, if the number of samples are
balanced with respect to each output category, c-index is equivalent
to the percentage of data points correctly classified by the algorithm
\citep{ripley96}. However, the latter measure is significantly biased
for unbalanced data sets, and c-index should be preferred in such
cases \citep{ripley96}. For the current study, we used c-index to
assess the performances of GLM and backpropagation networks and percent
of the samples correctly classified for k-NN, LDA, QDA, and ARTMAP
models. In all of the data sets considered in this study, whenever the
number of data points corresponding to each output category were unbalanced,
the data is truncated by randomly discarding necessary number of samples
corresponding to the output category with excessive number of samples.
This ensures the compatibility of different performance indices 
(c-index and percent correctly classified) in different
models. Hence, the data used in this study were perfectly balanced
with respect to the output categories.

\subsubsection{Overtraining, Cross-Validation and Importance of Independent Test}

As briefly mentioned in the introduction, the error rate of a given
model decreases monotonically toward zero on training data set as
the model is fitted to the training data. In other words, the model
loses its ability to generalize as it is fitted to the training data
more and more, a phenomenon termed as 'overfitting', or 'overtraining'
in statistical modeling literature \citep{bishop95,fielding99,hastie:etal01}.
To avoid overfitting of the data during training, a cross-validation
procedure may be employed during training \citep{bishop95,ripley96,hastie:etal01}.
For cross-validation, the training data set is shuffled and a certain
amount of data is used as a holdout subset at each iteration. At each
iteration, model is fit to the rest of the data and the predictive
performance is validated on holdout set, allowing the algorithm to
stop training at a point optimal to avoid overfitting and be able
to capture the overall characteristics if the system and not the peculiarities.

Bootstrapping of the training data also enables to determine the maximum
number of epochs to train the models, as well as to assess the stability
and variability of the model estimates \citep{bishop95}. To bootstrap
the model, $N$ draws with replacement are performed from the training
dataset, where $N$ is training set sample size, and this process
was repeated $K$ times, to create $K$ booted data sets, each of
size $N$. The booted data sets are samples of the original data set.
Generally, original data set is regarded as a sample of a larger universe
of data to which we wish to generalize \citep{hastie:etal01}. Thus,
this procedure allows to explore the behavior of a distribution of
booted models and permits to derive statistics and conclusions, which
may be applied approximately to the behavior of the original dataset
relative to the larger universe of data. For each booted model, the
number of epoch beyond which overfitting occurs is then determined
by standard methods \citep{bishop95,hastie:etal01} and the final
number of epochs to train the model with the original data set is
determined accordingly.

\subsection{Implementation Details}

All the models in our study have been trained using bootstrapping
and cross-validation to optimize the predictive power and generalizability
of the models. LDA, and QDA models were implemented in R-language
statistical software \citep{rlang}. GLM and backpropagation models
were implemented using NevProp3 software \citep{goodman96:software}.
k-NN and Fuzzy ARTMAP models are implemented in Matlab version
7 (Mathworks Inc.). Implementation details for GLM and backpropagation
models can be found for data sets 1 - 6 in \cite{tan:beklioglu2005},
\cite{tan:beklioglu:subm}, \cite{ozesmis99}, \cite{bahtiyar04}
and \cite{per2003}, respectively.

\section{Results}

\label{sec:results}

\begin{figure}
\begin{center}\includegraphics[%
  height=0.65\textheight,
  keepaspectratio]{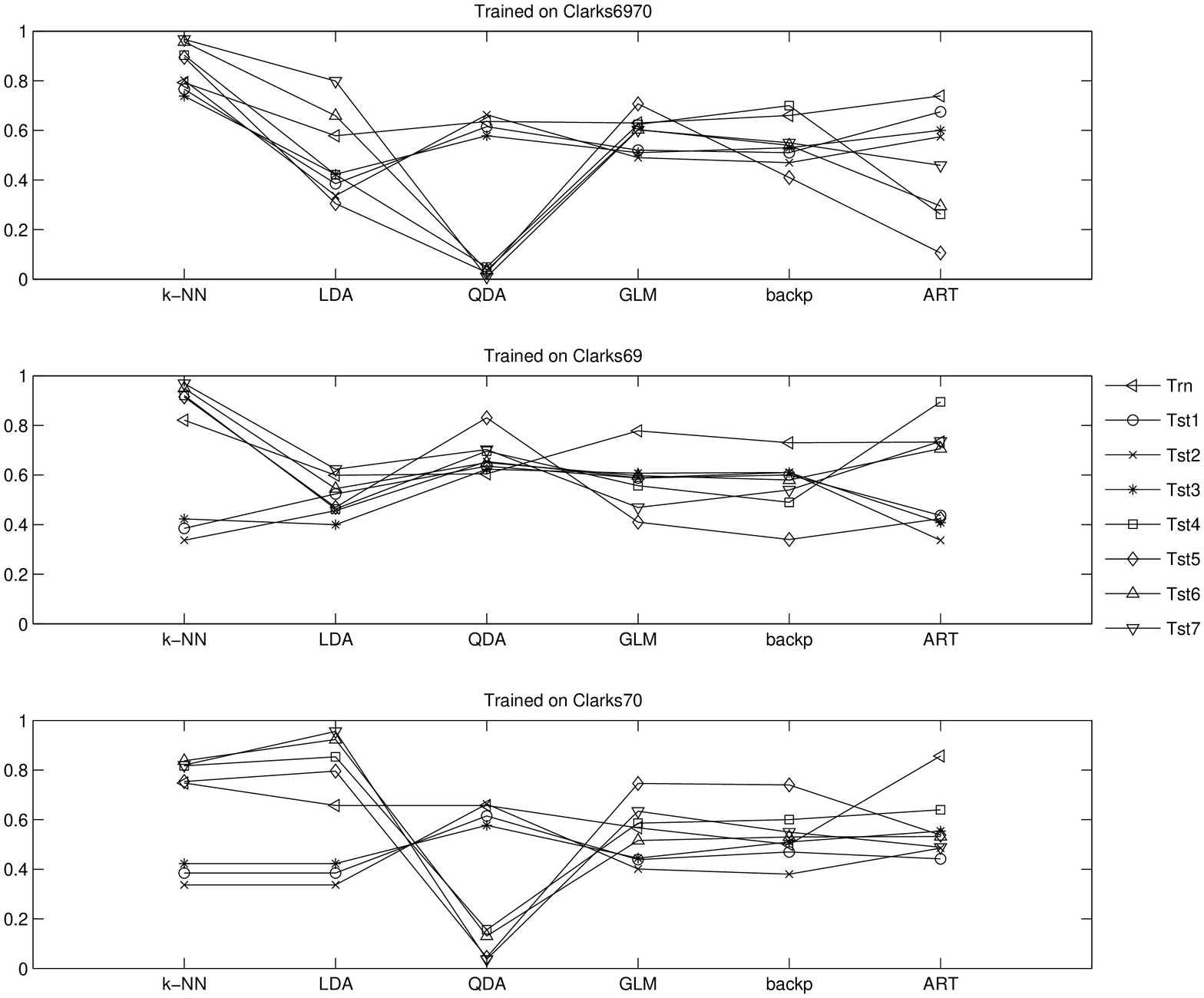}\caption{}\label{fig:robertson}\end{center}
\end{figure}

All the models were run with 10 different random seeds to assess the
variability of the estimates with initial conditions \citep{ozesmi:etalinpress}.
Moreover, all the models were run using 100 bootstraps and 5-fold
cross-validation with 10\% percent holdout to avoid overfitting. The
standard deviation of the performance assessment criteria (percent
correctly classified or c-index) were low and within $\pm5$\% of
the mean value, unless otherwise noted. Independent tests were run
using the random seed which achieved best performance on the training
data sets after cross-validation. Tables \ref{tab:eymir:mogan} -
\ref{tab:esra} summarizes the performances of k-NN, LDA, QDA, GLM,
backpropagation and ARTMAP models on both training and test samples
on different data sets considered in the study except for Data Set
4, for which the performances of the models are summarized in figure
\ref{fig:robertson}.

The performances of different models on \textbf{}Lakes Eymir (training
set) and Mogan \textbf{}(independent test set) (data set 1) are summarized
in Table \ref{tab:eymir:mogan}. All models constructed in this study
performed considerably well on this data set on training. 

The performances of all models were above 0.9 with the exception of
k-NN model, and was 1.00 with ARTMAP model, indicating that ARTMAP
classified all the data points in the training set correctly with
3 committed nodes. 

The backpropagation model was not trained for this particular data
set, since the high performance of GLM model rendered the computational
burden associated with backpropagation models in this case unnecessary
(see Discussion). However, despite their considerable success on the
training set, none of the models performed considerably better than
random (0.5) on test data, with the noticeable exception of GLM, which
predicted 82\% of the test cases correctly (Table \ref{tab:eymir:mogan}). 

\label{sec:res:5lakes}

Table \ref{tab:5lakes} summarizes the performances of different models
on central Anatolian shallow lakes (data set 2) \citep{tan:beklioglu:subm}.
As mentioned in section \ref{sec:5lakes}, this particular data set
included data from 5 different shallow lakes located in the same climatic
zone as training set, and part of this data is randomly excluded from
the training set and spared as an validation test set. Data from Lake
Mogan is also used as a second test set, which is spatially and temporally
distinct from training data. All of the models performed noticeably
good on this data set during training, the performances being close
to or above 0.8. However, note that backpropagation model reaches
a performance of 0.99 with 5 hidden units, while it takes ARTMAP 11
committed nodes to reach to the same level of performance (0.98; Table
\ref{tab:5lakes}).

If the validation test set was taken into consideration, all of the
models but QDA still performed better than random chance level (0.5) on data set 2,
with obvious superiority of the neural network based models (GLM,
backpropagation and ARTMAP). Among those, the difference in the validation
test performance was negligible. The difference, however, became apparent
on the performance on Lake Mogan test set, at which, despite of its
distinctiveness, connectionist approaches, GLM and backpropagation,
performed significantly better than ARTMAP model. Both GLM and backpropagation
models classified all 24 of the test cases correctly. Considering
the fact that ARTMAP uses a higher number of abstract category representations
(i.e., committed nodes) compared to the backpropagation model (number
of hidden units), apparent poor performance of ARTMAP on independent
test set is not surprising (see discussion).

The performances of different models on nest occurrence of red-winged
blackbird (data set 3) are summarized in Table \ref{tab:erie}. On
this particular data set, there is a clear dominance of k-NN, among
with ARTMAP, in terms of training set performance, over the traditional
models, LDA and QDA, and connectionist approaches, GLM and backpropagation.
k-NN and ARTMAP models' performance was 0.8, while other models' were
around 0.6 - 0.7 (Table \ref{tab:erie}). Considering the degrees
of freedom of k-NN and ARTMAP, which had higher performances in classifying
the training data, k-NN achieved a classification performance of 0.8
with 2 degrees of freedom, while ARTMAP required to use 14 degrees
of freedom (number of committed nodes; i.e., abstract category representations)
to achieve the same performance. However, none of the models performed
better than random on test data sets, the performance indices being
around 0.5, with a slight improvement in GLM and backpropagation models
on two test sets. All 6 of the methods failed to classify the test
sets which consist of the data collected from a spatially and/or temporally
distinct system (All Saints, Clarkes, Stubble and Darr marshes) effectively. 

The performances of different models on \textbf{}breeding presence
of red-winged black-bird (data set 4) \citep{robertson72} are summarized
in Figure \ref{fig:robertson}. Accordingly, on the training data sets, k-NN
performed considerably better than the other models, with a percent
correctly classified ratio of 0.8, followed closely by ARTMAP. A clear
exception is that ARTMAP performed close to 0.9 on the training set,
which consisted of the data collected only in 1970. On all three training
sets, nevertheless, k-NN and ARTMAP appeared to perform reasonably
well on predicting the breeding presence compared to other models.
On the other hand, inspection of the performances of these two models,
k-NN and ARTMAP, on independent test data sets reveals that k-NN predicts
the breeding success better than ARTMAP in almost all cases. Note
that the performance of k-NN models, which are trained on the data
collected either in 1969 or 1970, on test sets degrades, if the models
are tested on All Saints data collected during 1969 and/or 1970. On
the other hand, the performance is reasonably well on samples collected
from Stubble Patch and Darr marsh in 1995 and/or 1996. Nevertheless,
when trained on the complete data set (that is Clarkes marshes 1969
and 1970; train set 1), k-NN model successfully predicted
the breeding success in all of the test cases without any exceptions
(Figure \ref{fig:robertson}).

Table \ref{tab:bahtiyar} summarizes the performances of different
models on habitat selection data for \textit{L. senator},
\textit{C. brachydactyla} and \textit{H. pallida} (data set 5; \citealp{bahtiyar04}) on training
and independent test sets. All models were built
for each species by splitting the data into training and testing sets.
When the performances on training sets are considered, discriminant
analyses, both linear and quadratic, performed considerably worse
than other models. The performances of k-NN and GLM models are similar
in terms of training sets, while the performances of backpropagation
and ARTMAP are noteworthy, by classifying all of the training sets
correctly, with the exception of backpropagation model on the training
set of \textit{H. pallida}. However, note that in contrast with the
same models' performance on \textbf{}central Anatolian shallow lakes
(data set 2), ARTMAP achieved a performance of 1.00 with considerably
less number of abstract category representations (2-4 committed nodes)
than backpropagation models (5-8 hidden units), and thus ARTMAP in
this case was expected to be more generalizable than backpropagation
models. Not surprisingly, the performances of ARTMAP models independent
test sets are noteworthy, being close to 1 for each case, while backpropagation
models suffer from being close to random chance level on independent
test sets, with few exceptions.

Table \ref{tab:esra} summarizes the performances of different models
on habitat selection data for bird species in the central
Anatolia (data set 6; \citealp{per2003}) on training and test sets.
For this particular data, 9 sets of models have been built, 5 of which
are trained using the data of a particular species collected in Sultan
marshes and tested on the data of the same species collected in Lake
Tuzla (Sets 1-5 in Table \ref{tab:esra}; independent tests); and
remaining 4 sets consisted of the models trained on the half of the
data collected for a given species in Lake Tuzla, and tested in the
other half (Sets 6-9 in Table \ref{tab:esra}; validation). In general,
ARTMAP and k-NN models seem to have a superior predictive performance
on training data, followed by GLM and backpropagation models. QDA
models could not have been applied in 6 out of 9 cases for this data
set due to numerical instabilities and deficiencies in the data. Linear
and quadratic (where applicable) discriminant analyses appear to have
a predictive performance not better than random chance level. Comparison
of predictive performances on validations and independent tests indicates
that k-NN has a better predictive performance compared to that of
ARTMAP, successfully predicting the test data sets above chance level
with the only exception of Set 5.

\section{Discussion}

\label{sec:discussion}

\citet{seoane:etal2005} argued about the redundancy of independent
tests in predictive models in ecology, claiming that there is no particular
interest in estimating the predictive ability of a model in a universe
different from which it was built. However, the importance of the
ability of a given model, in terms of its performance on an independent
test data set, is emphasized in several studies \citep{ozesmis99,ozesmi:etalinpress,tan:beklioglu2005,tan:beklioglu:subm}.
Intuitively, observations from a given system corresponds not to a
universe of events, but rather to finite-size samples from a larger
universe of events. Hence, an ideal statistical model should be able
to predict not only the outcome of events (samples) on which it was
built, but also to predict the states of the system in the face of
events which has not been encountered in the finite size samples.
Hence, a given model should minimize the error rate on training set
(that is, samples used to fit the model) at the same time maximizing
the performance on an independent data set which has not been encountered
before, if the model is to be robust \citep{bishop95,ripley96,hastie:etal01}. Thus,
its performance on independent test data, shows its ability to generalize,
and indicates the robustness of the model for a given system, rather
than its performance on training set, which simply indicates its ability
to fit to the sample at hand.

The ability of a given model to avoid overfitting and to generalize
depends on how closely a model maps the input space to output space,
that is on the number of abstract category representations corresponding
to combinations of predictive vectors, in the case of 'global' models
such as GLM, backpropagation \citep{bishop95} and ARTMAP \citep{carpenter:etal1992}.
In the case of traditional classification models, it depends on the
number of neighborhoods for k-NN model, and on the number of output
classes and predictive variables for LDA and QDA \citep{hastie:etal01}.
Thus, the 'flexibility' of k-NN, LDA and QDA is fixed, equal to $k=2$
for k-NN, and is proportional to the number of predictive variables
for LDA, QDA and GLM, in our case. For backpropagation, it is represented
by the number of hidden units determined \textit{a priori}, and for
ARTMAP by the number of committed nodes after training. 

In the extreme case, the 'flexibility' of a model could be equal or
more than the number of observations (i.e., training points), and
in that case, the model would 'memorize' the training data, fitting
perfectly, while any observation different in the test samples from
the training points would cause a random prediction, hence rendering
the ability to generalize impossible. This corresponds to so called
'overfitting' of a model \citep{bishop95,fielding99,goodman96:manual}.
Thus, the ability of a given model to generalize on independent test
data would be evident from a trade-off between its performance on
training set and the flexibility of the model achieving that performance.
The importance of the flexibility of a model for fitting the training
data and of the independent test to assess the actual performance
of a given model is evident in our study as well.

\subsection{Data Sets 1 an 2: Lakes Eymir and Mogan, and Central Anatolian Shallow
Lakes}

In the case of data set 1, although ARTMAP performs better than other
models, independent test performance of GLM, which is considerably
better than ARTMAP, renders GLM to be applicable in that case. Even
a more drastic case took place \textbf{}in data set 2\textbf{,} on
which backpropagation model with 5 hidden units and ARTMAP model with
11 committed nodes predicted almost the same fraction of the training
sample. Their predictive performances on validation data set, which
consisted of the fraction of data initially split from the original
data, are also appear to be in close proximity. However, if the same
models were tested on an independent test, which consisted of data
collected from a spatially and temporally distinct system, predictive
performance of ARTMAP model dropped drastically to 0.67, whereas backpropagation
model retained its robustness. \textbf{}Considering the fact that
ARTMAP used a higher number of abstract category representations (i.e.,
committed nodes) compared to the backpropagation model (number of
hidden units), apparent poor performance of ARTMAP on independent
test set is not surprising. One can argue, however, that the system
constituting to the independent test sample might be governed be completely
different dynamics and as such it cannot be predicted by a model trained
on separate systems. However, all 5 lakes constituting to training
and test data sets are located in the same climatic zone, and all
these 5 lakes are ecologically governed by more or less similar mechanisms,
as far as the predictive variables concerned \citep{beklioglu:etal04,tan02,tan:beklioglu:subm}.
Furthermore, the fact that backpropagation model indeed predicted
all the cases in the test set correctly renders such an argument unlikely.

\subsection{Data Set 3: Nest Occurrence of Red-Winged Blackbird}

In the case of data set 3, k-NN and ARTMAP \textbf{}models appear
to have a greater predictive power on the training set compared to
other models, the performance criteria being around 0.8. However,
none of the models performed better than random on test data sets,
the performance indices being around 0.5, with a slight improvement
in GLM and backpropagation models on two test sets. This might be
because of the small sample sizes used as training set (230) though
this sample size is considerably larger than that of the data set
1 ($N=91$). Moreover, as mentioned in section \ref{sec:robertson:erie},
because one set of the models were developed to predict breeding success
and the other nest occurrence, the assumption has been made \textit{a
priori} that a high probability of nest occurrence corresponds to
a high probability of breeding success and vice versa. In addition,
since the Connecticut wetland variables did not include stem density,
the average value of stem density from the Lake Erie wetlands was
used when testing Connecticut wetlands data on the Lake Erie model.
Note also that a second \textit{a priori} assumption, namely that
stem height and nest height were correlated, was made for using these
two sets of data as independent tests for each other. For that reason,
it is quite likely that these \textit{a priori} assumptions have been
violated by the data, and further data collection, or further characterization
of habitat variables, might be required in order to ensure the compatibility
of these two sets with each other and to improve the predictive performances
of the models on both training and test sets.

Regardless of the underlying reason, however, this poor performance
of all the models on test sets emphasizes the importance of independent
test for assessing the actual predictive performance of a given model.
If, for example, ARTMAP model trained in the data collected in 1995
were applied for predicting the nest occurrence in the same area in
1996, based purely on its relatively high performance on training
data set, it would produce misleading results. This is indicated by
the fact that its performance on test data collected in the same area
in 1996 is not better than random chance level. In this case, it is
apparent that the model should be improved by, for example, obtaining
more samples, or changing the model structure and/or type. Model improvement
is beyond the scope of the current study. Nevertheless, there are
several techniques based on information theoretical approaches for
improving the predictive performance of models, readily available
in the literature (e.g., \citealp{hastie:etal01}), and particular
examples for predicting habitat selection and distribution of bird
species are provided in the literature for the case of general additive
models (GAMs), which are marginally related to GLMs \citep{bustamante:seoane2004,seoane2004a,seoane:etal2004b}.

\subsection{Data Set 4: Breeding Success of Red-Winged Blackbird}

For data set 4, a general pattern emerged for all three sets that
k-NN and ARTMAP models are again superior to other techniques on the
training set performance. Inspection of the independent test results,
however, revealed that, k-NN has a broader ability to generalize over
the new data sets. \textbf{}It is especially intriguing that the performance
of k-NN models, which are trained on the data collected either in
1969 or 1970, on test sets degrades when the models were tested on
All Saints data collected during the same period, while the performance
is reasonably well on samples collected in Stubble and Darr marshes
collected in 1995 and 1996. Nevertheless, as apparent from Table Figure
\ref{fig:robertson}, when trained on the complete data set (train
set 1), k-NN model successfully predicts the breeding success in all
of the test cases without any exceptions. This might probably be as
a result of the fact that the data in this case covers a relatively
broad temporal domain (2 years instead of 1 year). \textbf{}Note also
that in the case of Robertson data, \textbf{}performance on the independent
tests are boosted for spatial sets, that is, the models are in general
more generalizable on temporal domain but as such is not true for
spatial domain, this might in turn indicate the importance of training
data set.

\subsection{Data Set 5: Breeding Presence of Three Bird Species}

For data set 5, there was a clear dominance of ARTMAP models, in terms
of the predictive power on both training and test sets, Note that
in contrast to the same models' performance on central Anatolian shallow
lakes (data set 2), ARTMAP achieved a performance of 1.00 with considerably
less number of abstract category representations than backpropagation
models. Thus ARTMAP is expected to be more generalizable than backpropagation
models. For this case, when the performances on training sets are
considered, discriminant analyses, both linear and quadratic, performed
considerably worse than other models. Surprisingly, the performances
of ARTMAP models on independent test sets is noteworthy, being close
to 1 for each case, while backpropagation models suffer from being
close to random chance level on validation and independent test sets,
with few exceptions.

\subsection{Data Set 6: Habitat Selection of Bird Species in Central Anatolia}

When considering data set 6, it was apparent that again, k-NN and
ARTMAP models are superior to other models, in terms of their predictive
performance on training data sets. Nevertheless, comparison of predictive
performances on validation and independent tests indicated that k-NN
had a better predictive performance compared to that of ARTMAP. k-NN
successfully predicted the test data sets above chance level with
the only exception of Set 5 . Note that in the case of these data
, sets 1-5 include test samples collected from a spatially distinct
region than the samples for training data. This , in turn, indicates
that k-NN models, in this case are particularly robust in terms of
their ability to generalize over new data sets.

\subsection{General Discussion}

On all data sets, traditional discriminant analyses, linear and quadratic,
had a poor predictive performance on both training and independent
test data sets. Note, however, from section \ref{sec:traditional}
that discriminant analyses strictly require that the underlying data
to be sampled from a Gaussian distribution. Several studies in the
literature appear to employ discriminant analysis to infer species
distribution or occurrence, and attain reasonably good predictive
performance \citep{joy:death2003,maron:lill2004}, in contrast to
our study. Despite their considerable success, however, we do not
recommend traditional discriminant analysis for prediction purposes,
unless one makes sure that the underlying data is distributed appropriately,
or filtered through a suitable transformation in preprocessing stage
to satisfy the required parametric distribution.

Statistical learning models associate a probability with each alternative
state given the simultaneous observation of all variables at a given
time step (or the whole set of past observations in the case of unsupervised
methods) \citep{kosko1992}. Once the model is trained, resulting
probability densities associated with each state are used to predict
and forecast the state the system will occupy based on the new observations.
For predictive modeling of ecosystems which are known to exhibit multiple
stable states and catastrophic regime changes \citep{may76,scheffer:etal93:ASS},
this probabilistic design can also be exploited as (at least) a qualitative
measure of the distance to threshold for regime changes, in addition
to identifying bifurcations and regime shifts in ecosystem dynamics,
by combining probabilities associated with the points in input space
with the sensitivity analyses, which systematically scan the input
space \citep{recknagel:etal97,scardi01,tan:beklioglu2005}. Presentation
and elaboration of sensitivity analyses are beyond the scope of this
study, however, sensitivity analyses of GLM and/or connectionist neural
network models for our data sets can be found elsewhere \citep{per2003,bahtiyar04,tan:beklioglu2005,tan:beklioglu:subm,ozesmi:etalinpress},
and a through discussion of their use in identifying regime shifts
and thresholds associated with these shifts for data sets 1 and 2
(Lakes Eymir and Mogan, and central Anatolian shallow lakes) can be
found in \citet{tan:beklioglu2005} and \citet{tan:beklioglu:subm},
respectively.

Note that two main types of ecological data have been considered in
this study: one type of data (data sets 1 and 2) consisted of time-dependent
data, meaning that the dynamics governing these systems depend on
the passage of time, including small- to large-scale periodicities.
However, the second type (data sets 3-6) was time-independent, or
strictly stationary, in the sense that possible periodic dynamics and trends
governing these systems are expected to be well beyond the time-scale
of the data collection. Statistical classification methods are, however,
known to be insensitive to the time correlations or interactions of
individual variables over a time series, i.e they can not capture
temporal system dynamics, unless an explicit independent variable
reflecting the time-dependence and periodicities within the series
is included \textit{a priori,} as such is also true for connectionist
neural network models.

\subsubsection{Time-Dependency of the Data and Selection of Suitable Model}

Note that on the first type of data (data sets 1 and 2), connectionist
neural network models (GLM and multilayered feedforward backpropagation)
achieved a fairly good predictive performance, both on training and
test data sets. These models apparently captured the nonlinear interactions
between the variables as well as inherent non-stationarity of the
data. However, note that for one of these data (data set 1), we included
the z-score transformation of the water level, which inherently reflects
the periodic changes due to the seasonal periodicities of the system.
In the other data set (2), we included an explicit representation
of the time series (period index) as well as water level data which,
again, was expected to reflect the periodicities inherent in the system.
\citet{meggs:etal2004} used a generalized linear modeling approach
to predict the occurrence of a lucanid beetle species based on habitat
variables, but their model attained to a relatively moderate predictive
performance on predicting species abundance and occurrence. We suggest
that the moderate discriminatory ability of their model is, at least
partly, mediated by the lack of account to periodic dynamics. Similarly,
\citet{dunk:etal2004} employed a generalized additive model, which
is analogous to generalized linear models, to predict the occurrence
of mollusk species, and showed that the inclusion of climatic variables
contributed significantly to the predictive ability of their model.
This is presumably because climatic variables reflect at least seasonal
periodic dynamics of the system, considering that climatic forces
are among the most important driving forces for a given ecosystem.
As connectionist neural network model are closely related to generalized
linear models (section \ref{sec:glm}), the same reasoning is applicable
to these models as well. Thus, we suggest that for data sets over
time-scales which are smaller than the seasonal and other possible
periodic dynamics of the system, an explicit predictive variable should
be included in the model that potentially reflects these periodic
dynamics, either directly (as period index), or indirectly (as water
level z-score), at least for the case of GLM or connectionist neural
network models.

Existing literature of predictive modeling in ecology emphasizes the
advantages and predictive power of connectionist neural network approaches
owing to these models' inherent ability to capture complex nonlinear
interactions between the predictive variables \citep{lek:etal96,lek:guegan99,scardi96,scardi01}.
However, this celebrated predictive performance also brings about
an increased complexity and thus an increased computational costs
resulted from this complexity. The computational cost of these models
could be overwhelming especially with larger data sets. However, as
the model on data set 1 shows, GLMs could attain a significant predictive
performance, at the same time avoiding the costs of neural network
models. Thus, we suggest that one should fit a GLM to the data set
before considering a neural network approach, a high predictive power
of which would render the computational cost associated with neural
networks unnecessary. 

For the time-independent data sets (data sets 3-6), our results show
that neighborhood-based methods, k-NN and ARTMAP, are superior in
terms of their predictive performances compared to other techniques,
both on training and test data sets. Test \textbf{}performance of
k-NN and ARTMAP models on data set 3 appears to be an exception to
this, and possible reason for this exception is discussed above. Based
on our results, nevertheless, we suggest that k-NN and ARTMAP models
are more suitable for spatial data, such as habitat selection and
species distribution instead of more dynamic alternatives such as
GLM and connectionist neural network models, at least if the time-scale
considered is assured to be relatively insignificant compared to the
time-dependent periodicities governing ecosystem dynamics. 

ARTMAP, although considered as a neural network architecture, is implicitly
a neighborhood-based classification technique. A geometric interpretation
of the operation of ARTMAP \citep{carpenter:etal1991c,carpenter:etal1992}
suggests that its operation is analogous to k-NN neighbor method.
An important difference is that while the size of the neighborhood
is fixed for k-NN, it is adaptive in ARTMAP, adjusted on-the-fly depending
on the performance of the model on current data point (and previous
ones). Thus, analogous to the relation between GLM and connectionist
neural network models, we consider k-NN to be a relatively primitive,
computationally less expensive alternative to ARTMAP. On our spatial
data sets, k-NN performed considerably better than ARTMAP on 2 of
the 3 cases (data sets 4 and 6) on independent test sets. Hence, we
suggest that k-NN should be considered for predictive spatial modeling
before considering a more advanced but complex model such as ARTMAP.

\section{Conclusions}

Our study suggests that different methods for statistical predictive
modeling of ecosystems are suitable, depending on the data sets and
ecosystem dynamics that are to be modeled. For the cases involving
data sets whose underlying distribution is unknown, or presumed to
be irregular, traditional statistical models such as discriminant
analyses have poor predictive performances and thus could lead to
misleading and invalid predictions. For the data sets involving time-dependent
dynamics and periodicities whose frequency are possibly less than
the time scale of the data considered, GLM and connectionist neural
network models, such as multilayer feedforward backpropagation models,
appear to be most suitable, in terms of their performance on both
training and test sets, provided that a predictive variable reflecting
these time-dependent dynamics, either implicitly or explicitly in
included in the model. For spatial data, which does not include any
time-dependence comparable to the time scale covered by the data,
on the other hand, neighborhood based methods such as k-NN and ARTMAP
proved to be more robust than other methods considered in this study.
However, for predictive modeling purposes, one should consider applying
first a suitable, computationally inexpensive method to the data at
hand, a good predictive performance of which would render the computational
cost and efforts associated with complex variants unnecessary. Further
characterization of the data included in this study using different
and/or variants of the methods considered here, as well as application
of the models considered here to new data sets would, nevertheless,
reveal further characterizations and suggestions for suitability and
applicability of statistical predictive modeling techniques in ecology.

\section*{Acknowledgments}

This study is partly supported by TUBITAK grant YDABAG-100Y112 and
partly by research grant provided by Middle East Technical University.
We thank Raleigh J. Robertson for kindly providing the data for breeding
success of bird species in Clarkes Pond, and Hillary Welch and Geoff
Welch for providing the data for Data Set 5. Stacy \"{O}zesmi, Ay\c{s}eg\"{u}l
Doma\c{c}, \"{O}nder Cirik, Zeren G\"{u}rkan, and Paul Hope are
thanked for their help during data collection and bird census. Stacy
\"{O}zesmi is also thanked for her help in modeling Data Sets 3 and
4. We also thank Mert Alt\i nta\c{s} for his help with preprocessing
satellite imagery and GIS maps, and C\"{u}neyt Kuban\c{c} for his
helpful comments on backpropagation and GLM models of Data Set 5.

 \newpage


\section*{Tables}

\begin{table}

\caption{Models trained on Lake Eymir, tested on Lake Mogan data \citep{tan:beklioglu2005}.
\textit{N}: number of data points; \textit{P}: number of independent
variables. k-NN, LDA QDA and ARTMAP results are given as percent correctly
classified, backprop and GLM as c-index (corrected c-index for training
set). Integers indicated before the performance values of the training
sets for ARTMAP model indicate the number of
committed nodes.}

\begin{center}\begin{tabular}{l|cccccccc} \textbf{Set}&\textbf{N}&\textbf{P}&\textbf{k-NN}&\textbf{LDA}&\textbf{QDA}&\textbf{GLM}&\textbf{BackProp}&\textbf{ARTMAP}\\ \hline Training&91&5&.846&.939&.969&.963&---&3;.1.000\\ Independent Test&43&5&.429&.524&.476&.815&---&.643\\ \end{tabular}\end{center}

\label{tab:eymir:mogan}
\end{table}

\newpage

\begin{table}

\caption{Models trained on Anatolian Lakes, tested on validation test set,
which consisted of data randomly split from training set and did not
included in model fitting phase (see text) and Lake Mogan data \citep{tan:beklioglu:subm}
as independent test. \textit{N}: number of data points; \textit{P}:
number of independent variables k-NN, LDA QDA and ARTMAP results are
given as percent correctly classified, backprop and GLM as c-index
(corrected c-index for training set). Integers indicated before the
performance values of the training sets for backpropagation and ARTMAP
models indicate the number of hidden units and number of committed
nodes, respectively, of backpropagation and ARTMAP models.}

\begin{center}\begin{tabular}{l|cccccccc} \textbf{Set}&\textbf{N}&\textbf{P}&\textbf{k-NN}&\textbf{LDA}&\textbf{QDA}&\textbf{GLM}&\textbf{BackProp}&\textbf{ARTMAP}\\ \hline Training&440&5&.998&.814&.773&.943&5;.986&11;.977\\ Validation&101&5&.816&.802&.255&.962&.998&.956\\ Independent Test&24&5&.750&1.00&.833&1.00&1.00&.667\\ \end{tabular}\end{center}

\label{tab:5lakes}
\end{table}

\newpage

\begin{table}

\caption{Models trained on Lake Erie, tested on Lake Erie, All Saints and
Clarkes marshes data \citep{robertson72,ozesmi96,ozesmi:mitsch97}.
\textit{N}: number of data points; \textit{P}: number of independent
variables. k-NN, LDA QDA and ARTMAP results are given as percent correctly
classified, backprop and GLM as c-index (corrected c-index for training
set). Integers indicated before the performance values of the training
sets for backpropagation and ARTMAP models indicate the number of
hidden units and number of committed nodes, respectively, of backpropagation
and ARTMAP models.}

\begin{center}\begin{tabular}{l|cccccccc} \textbf{Set}&\textbf{N}&\textbf{P}&\textbf{k-NN}&\textbf{LDA}&\textbf{QDA}&\textbf{GLM}&\textbf{BackProp}&\textbf{ARTMAP}\\ \hline Training (s95d95)&230&6&.822&.648&.644&.716&6;.730&14;.826\\ Independent Test-1 (s96)&98&6&.541&.592&.591&.681&.670&.459\\ Independent Test-2 (d96)&84&6&.560&.524&.536&.578&.550&.440\\ Independent Test-3 (AllSaints69)&68&6&.500&.501&.500&.380&.430&.382\\ Independent Test-4 (AllSaints70)&110&6&.501&.500&.501&.470&.520&.518\\ Independent Test-5 (Clarks69)&124&6&.516&.540&.589&.660&.660&.540\\ Independent Test-6 (Clarks70)&108&6&.472&.444&.435&.480&.470&.454\\ \end{tabular} \end{center}

\label{tab:erie}
\end{table}

\newpage

\begin{table}

\caption{Models trained and tested on bird habitat selection data\citep{bahtiyar04,welch2004}.
\textit{N}: number of data points; \textit{P}: number of independent
variables. k-NN, LDA QDA and ARTMAP results are given as percent correctly
classified, backprop and GLM as c-index (corrected c-index for training
set). Integers indicated before the performance values of the training
sets for backpropagation and ARTMAP models indicate the number of
hidden units and number of committed nodes, respectively, of backpropagation
and ARTMAP models.}

\begin{center}\begin{tabular}{l|cccccccc} Training-1 (\textit{L. senator}) &274&12&.828&.781&.799&.859&8;1.00&2;1.00\\ Independent Test-1 (\textit{L. senator}) &273&12&.678&.780&.798&.781&.831&.971\\ \hline Training-2 (\textit{H. pallida})&246&12&.866&.488&.496&.759&3;.874&4;1.00\\ Independent Test-2 (\textit{H. pallida}) &245&12&.669&.486&.502&.703&.657&.980\\ \hline Training-3 (\textit{C. brachydactyla}) &294&12&.847&.646&.701&.855&10;1.00&3;1.00\\ Independent Test-3 (\textit{C. brachydactyla})&293&12&.765&.648&.703&.769&.809&.962\\ \end{tabular}\end{center}

\label{tab:bahtiyar}
\end{table}

\newpage

\begin{table}

\caption{Models trained and tested on bird habitat selection data \citep{per2003}.
\textit{N}: number of data points; \textit{P}: number of independent
variables. k-NN, LDA QDA and ARTMAP results are given as percent correctly
classified, backprop and GLM as c-index (corrected c-index for training
set). Integers indicated before the performance values of the training
sets for backpropagation and ARTMAP models indicate the number of
hidden units and number of committed nodes, respectively, of backpropagation
and ARTMAP models. k-NN model outperforms other models on both training
and test performance, with the exception of independent test-5 (highlighted).
Although ARTMAP seem to have a high performance on training data,
its performance is lower than GLM and Backprop on independent tests,
being at random chance level for four of the independent tests (highlighted).}

\begin{center}\begin{tabular}{l|cccccccc} \textbf{Set}&\textbf{N}&\textbf{P}&\textbf{k-NN}&\textbf{LDA}&\textbf{QDA}&\textbf{GLM}&\textbf{BackProp}&\textbf{ARTMAP}\\ \hline 

Training-1 (ss-acraru)&74&12&.973&.905&---&.959&---&4;.946\\ 

Independent Test-1 (tuzla-acraru)&506&12&.968&.945&---&.936&---&.986\\ \hline 

Training-2 (ss-alaarv)&74&12&.986&.824&.851&.807&---&11;960\\ 

Independent Test-2 (tuzla-alaarv)&505&12&.798&.430&.551&.705&---&\textbf{.412}\\ \hline 

Training-3 (ss-calruf)&118&12&.992&.788&---&.822&12;.925&5;.924\\ 

Independent Test-3 (tuzla-calruf)&506&12&.773&.530&---&.875&.909&.714\\ \hline 

Training-4 (ss-ciraer)&48&12&.917&.833&.896&.757&12;.905&8;.930\\ 

Independent Test-4 (tuzla-ciraer)&506&12&.915&.785&.332&.760&.829&\textbf{.453}\\ \hline 

Training-5 (ss-motfla)&50&12&.920&.760&.960&.696&12;.584&9;.960\\ 

Independent Test-5 (tuzla-motfla)&505&12&\textbf{.535}&.570&.941&.508&.683&\textbf{.521}\\ \hline 

Training-6 (tuzla-calbra)&50&12&.940&.780&---&.745&12;.900&4;.920\\ 

Validation-1 (tuzla-calbra)&516&12&.880&.629&---&.778&b.641&.800\\ \hline

Training-7 (tuzla-melcal)&118&12&.983&.729&---&.651&12;.902&6;.831\\ 

Validation-2 (tuzla-melcal)&515&12&.937&.604&---&.608&.657&\textbf{.555}\\ \hline

Training-8 (tuzla-milcal)&56&12&.982&.768&---&.698&2;.708&8;.946\\ 

Validation-3 (tuzla-milcal)&516&12&.856&.613&---&.676&.850&.785\\ \hline

Training-4 (tuzla-oenisa)&102&12&.990&.745&---&.858&2;.719&7;.892\\ 

Validation-4 (tuzla-oenisa)&505&12&.954&.529&---&.818&.835&.745\\ 

\end{tabular}\end{center}

\label{tab:esra}
\end{table}

\clearpage

\section*{Figure Captions}

\begin{description}
\item [Figure~1:]Schematic representation of fuzzy ARTMAP architecture.
Input vectors are processed in ART$_{a}$ module while target categories
are processed in ART$_{b}$ module. Semi-disks represent adaptive
weights. For details, see text (modified from \cite{carpenter:etal1992}). 
\item [Figure~2:]Predictive performance of models on breeding success
data of red-winged black-bird (data set 4). k-NN: k-nearest neighbor;
LDA: linear discriminant analysis; QDA: quadratic discriminant analysis;
GLM: generalized linear model; backprop: multilayer feedforward backpropagation
neural network. Each panel shows the training and test performances
of the models trained on Clarkes data. Upper panel: models trained
on Clarkes 1960-1970 data; Middle panel: models trained on Clarkes
1969 data; Lower panel: models trained on Clarkes 1970 data. trn: training performance; tst: test performance (see text). 
\end{description}
\clearpage


%

\newpage

\newpage

\end{document}